\documentclass[prb,showpacs,twocolumn]{revtex4}
\usepackage{graphicx}
\usepackage{dcolumn}
\usepackage{bm}
\begin{document}
\title{Reply to ``Comment on `Magnetic field effects on neutron diffraction 
in the antiferromagnetic phase of $UPt_3$'\,'' }
\author{Juana Moreno}
\altaffiliation[Present address: ]{Department of Physics and Astronomy, 
                Clemson University, Clemson SC 29634}
\author{J. A. Sauls}  
\affiliation{Department of Physics and Astronomy, Northwestern University, 
Evanston IL 60208}  
\date{\today}  
\begin{abstract} 
F{\aa}k, van Dijk and Wills (FDW) question our interpretation of elastic 
neutron-scattering 
experiments in the antiferromagnetic phase of UPt$_3$. They state that 
our analysis is 
incorrect because we average over magnetic structures that are disallowed 
by symmetry.
We disagree with FDW and reply to their criticism below. FDW also point 
out that we 
have mistaken the magnetic field direction in the experiment reported in 
Ref. \onlinecite{dij98}. We correct this error and note that our previous 
conclusion is
also valid for the correct field orientation.
\end{abstract}  
\pacs{74.70.Tx,75.20.Hr,75.25.+z} 
\maketitle

We disagree with the claim of F{\aa}k, et al.\cite{fak02} that our 
analysis of elastic neutron-scattering experiments in the antiferromagnetic 
(AFM) phase of UPt$_3$ is incorrect because we average over magnetic structures 
that belong to different 
irreducible representations of the crystallographic space group. 
Classification of magnetic 
structures and magnetic phase transitions on the basis of irreducible 
representations of 
the space group and time-inversion neglects the {\em fundamental role} that
exchange interactions play in magnetic phase transitions.
\cite{dzy64,and76,and80,toledano87}
Exchange interactions are invariant under continuous rotations of 
all the moments, and typically dominate the anisotropy energies that couple 
the atomic moments to the lattice. 
Classification of magnetic structures based on the exchange group 
accounts for the wide
variety of magnetic structures that are observed in magnetic materials.
The Shubnikov classification, which does not take into account the higher 
symmetry of the exchange interactions, disallows some
of these structures.\cite{and80}

Thus, for a magnetic instability driven by exchange interactions the primary 
irreducible representation is based on the combined group of continuous 
rotations in spin space, 
the crystallographic space-group and time-reversal, $\mathsf{G}_{\text{ex}}$.
The irreducible representations of the exchange group combine several 
irreducible representations of the space group.\cite{Izy91} 
Thus, not only are magnetic structures corresponding to irreducible 
representations of the space group allowed. On the contrary, structures 
that are a combination of irreducible representations of the space group, 
but belong to one exchange representation, are also possible magnetic 
structures. 
Many examples of magnetic structures with these type of 
``mixed space-group representations'' \cite{mixed}
are described in the literature.\cite{oles76,toledano87}

In most materials the magnetically ordered phase is defined by 
one irreducible representation of the space group due to the 
anisotropy energies which resolve (at least partially) orientational 
degeneracies within the exchange representation. \cite{anisotropy}
However, since the anisotropy terms are relatively weak, the energy splitting of 
differently oriented magnetic states are small. Thus, magnetic domain 
structures, including their response to magnetic fields,
should be analyzed using the degenerate, or nearly degenerate, states within 
the full exchange multiplet. We believe this is the correct approach to 
understanding the magnetism and to analyze the possible magnetic 
structures in the heavy fermion compound UPt$_3$.

In our analysis, we considered a general model for UPt$_3$ 
compatible with the available data.\cite{mor01}
We selected one irreducible representation of $\mathsf{G}_{\text{ex}}$ 
that is consistent with elastic neutron scattering data in zero field. 
If we neglect the spin-lattice couplings then only the relative orientations 
of the atomic moments in the magnetic unit cell are fixed by the 
primary irreducible representation. Anisotropy energies are also included
to resolve, or partially resolve, the degeneracies of the exchange 
representation.

Neutron scattering and X-ray experiments in UPt$_3$ show AFM 
order with propagation vector
$\vec{q}_1=\vec{a}^*_1/2$.\cite{3arms}
The magnetic U ions occupy two symmetry equivalent positions in the unit cell. 
The magnetic representation 
has 6 dimensions (3 times the number of magnetic ions). Until very recently, 
the crystal structure of 
UPt$_3$ was thought to be hexagonal with space group $D^4_{6h}$. However, a 
recent X-ray diffraction
experiment revealed a lower trigonal symmetry with space group $D^3_{3d}$.
\cite{wal01} In either case, 
the magnetic representation can be decomposed in six one-dimensional 
representations. Three of these 
correspond to FM alignment of the ions in the unit cell; the other three 
representations 
correspond to AFM alignments. The alignment of the magnetization or 
sub-lattice magnetization may be along 
the $\hat{x}$, $\hat{y}$ or $\hat{z}$ axes. 
However, these six structures are connected with only two exchange 
representations corresponding to FM or AFM alignment in the unit cell. 
Table \ref{Irreps} shows the irreducible 
representations and basis functions of the crystallographic space groups 
$D^4_{6h}$ and $D^3_{3d}$ grouped 
by their corresponding exchange multiplets.

\begin{table}
\begin{minipage}{0.9\hsize}
\begin{ruledtabular}
\begin{tabular}{|l|c|c||l|c|c|}
	   & FM              & AFM                &            & FM               & AFM				\\
\hline
           &$\tau_2\,:\,\hat{x}$ & $\tau_7\,:\,\hat{x}$ &            &$\tau_2\,:\,\hat{x}$ & $\tau_3\,:\,\hat{x}$		\\
$D^4_{6h}$ &$\tau_4\,:\,\hat{y}$ & $\tau_5\,:\,\hat{y}$ & $D^3_{3d}$ &$\tau_4\,:\,\hat{y}$ & $\tau_1\,:\,\hat{y}$		\\
	   &$\tau_6\,:\,\hat{z}$ & $\tau_3\,:\,\hat{z}$ &            &$\tau_2^{'}\,:\,\hat{z}$ & $\tau_3^{'}\,:\,\hat{z}$	\\
\end{tabular}
\end{ruledtabular}
\medskip
\caption[]{Irreducible representations and basis functions of the space groups $D^4_{6h}$ and $D^3_{3d}$ grouped 
by exchange multiplets. FM and AFM refer to ferromagnetic or antiferromagnetic alignment of the two U ions on 
the unit cell. We use the notation of Kovalev in Ref. \onlinecite{kovalev65}.}
\label{Irreps}
\end{minipage}
\end{table}

Our study is based on a free energy functional (Eq. 9 of Ref. 
\onlinecite{mor01}) which includes 
the exchange, anisotropy  and Zeeman energies.
First, a uniaxial anisotropy term 
(not shown in Eq. 9 of Ref. \onlinecite{mor01})
restricts the order parameter to the basal plane. In addition, 
the in-plane (hexagonal) anisotropy 
energy favors alignment of the moments along any of the three directions 
perpendicular to the hexagonal  lattice vectors. 
Note that the form of the anisotropy energy is the same for 
either $D^4_{6h}$ and $D^3_{3d}$ symmetry groups. The effect of a 
magnetic field on the AFM order is included
through the Zeeman coupling to the atomic moments, which in general mixes 
different nearly degenerate representations of the space group within the 
exchange multiplet.\cite{Moriya}
The competition between the anisotropy energy and the Zeeman coupling induces 
hexagonal modulations of the upper critical field as 
a function of the orientation of the field in the basal plane 
at the transition to the superconducting phase.\cite{kel94,sau96a} 
The in-plane anisotropy energy is small, since a large 
in-plane anisotropy energy would produce an orthorhombic modulation of the 
upper critical field, which is not observed. 
Higher order anisotropy terms \cite{example}
which might resolve the remaining 
degeneracy and thus favor alignment of the moments along the 
propagation vector of the magnetic 
order would be extremely small. Therefore, the three structures shown in 
Fig. 1 of Ref. \onlinecite{mor01} 
are degenerate, or quasi-degenerate, and certainly should be considered 
in the analysis of the magnetic 
structure and neutron scattering in the presence of an in-plane magnetic field.
Thus, in our analysis we consider the possibility of degenerate, or nearly 
degenerate, magnetic structures
by making an average over different distributions of domains. We also 
presented results 
and predictions for the single magnetic structure with the magnetization 
parallel to the propagation vector. 
The authors of the comment seem to have overlooked this prediction, which 
if we had confined our analysis
to a single representation of the space group, as F{\aa}k, et al. advocate, 
would be the only relevant structure.

We did mistake the magnetic field direction in the experiment reported in 
Ref. \onlinecite{dij98}.
In the correct geometry of that experiment the field was along the 
reciprocal lattice direction [-1,2,0].
The ratios reported in Eq. 5 of Ref. \onlinecite{mor01}, and in the 
paragraph that follows that equation,
should be modified as follows. When only domain ``1'' is populated we have 
$r=1$. For a crystal with equally populated magnetic domains, the correct 
ratio between the scattering rate at high field and zero field is
\begin{eqnarray}
r=\frac{1-(0.441\,\hspace{0.01in}\cos(\theta_H+\pi/2))^2}
{\langle 1-(0.441\,\hspace{0.01in}\cos(\theta))^2\rangle}
 =0.89
\,.
\end{eqnarray}
Our previous conclusion, stated for the incorrect field orientation, 
is unchanged for the correct field orientation; 
it is not possible based on existing data to conclude whether or 
not the U moments rotate with 
the field, because of the small change in intensity that is expected 
for this Bragg peak and the large error
bars that are reported for the intensity. We also concluded that, 
in order to understand UPt$_3$ 
magnetism in the presence of magnetic field or under pressure, 
systematic, zero-field measurements 
of the intensity of a number of magnetic peaks in the same single crystal, 
such as those
reported in Ref. \onlinecite{gol86}, need to be carried out.
Furthermore, our hypothesis that intrinsic stacking faults pin 
the AFM domain walls in the ab-plane 
and fix the spatial distribution of domains with different 
propagation vectors has been recently reinforced. 
For uniaxial pressures applied to the basal plane a significant 
increase in the magnetic intensity has been 
reported \cite{dij01} in contrast with the relatively small change 
in a magnetic field.\cite{lus96a,dij98}
Pinning by intrinsic stacking faults may help explain this difference, 
since the applied magnetic 
field leaves the distribution of regions with different propagation 
vectors unaltered. However, uniaxial 
pressure likely disturbs the configuration of stacking faults 
leading to a stronger effect on the magnetic 
structure.

In conclusion, our analysis of the neutron scattering data is based on a 
sound theoretical 
model for possible magnetic structures in UPt$_3$, which is more general than 
would be allowed based on a single irreducible representation of the 
space group. The relative importance of exchange interactions leads 
naturally to mixed irreducible representations of the crystal space group, 
which are relevant because they are energetically allowed.

\bibliographystyle{apsrev}

\begin{thebibliography}{16}
\bibitem{mixed} For some materials, such as the orthoferrites, 
approximately 20\% of the magnetic phases are described by multiple 
irreducible representations of the space group. \cite{oles76}
\bibitem{anisotropy} Anisotropy energies, which arise from 
spin-orbit, dipolar and indirect interactions, may also lead to mixing of 
magnetic structures belonging 
to different irreducible representations of $\mathsf{G}_{\text{ex}}$.
\bibitem{3arms} The propagation vectors 
$\vec{q}_2=\vec{a}^*_2/2$ 
and $\vec{q}_3=(\vec{a}^*_1-\vec{a}^*_2)/2$ are also present since the 
wave vector has three arms.
\bibitem{Moriya} Anisotropy energies can also 
couple an AFM representation with the FM order parameter belonging to a 
different exchange multiplet.
This leads to weak, Dzyloshinski-Moriya ferromagnetism. Since the induced 
magnetization couples linearly
to the external field we included this correction in our analysis, in 
addition to the quadratic Zeeman coupling.
\cite{mor01}.
\end{thebibliography}

\end{document}